  \documentclass[11pt]{article}

  \usepackage{latexsym}
  \usepackage{amssymb}
  \usepackage{amsmath}
  \usepackage{graphicx}

  \makeatletter
  \def\section{\@startsection {section}{1}{\z@}{-3.5ex plus -1ex minus
   -.2ex}{2.3ex plus .2ex}{\large\bf}}
  \def\subsection{\@startsection{subsection}{2}{\z@}{-3.25ex plus -1ex minus
   -.2ex}{1.5ex plus .2ex}{\normalsize\bf}}
  \makeatother

  \makeatletter
  \@addtoreset{equation}{section}
  
  \makeatother

  \newcommand{\nc}{\newcommand}

\def\slash#1{\setbox0=\hbox{$#1$}#1\hskip-\wd0\hbox to\wd0{\hss\sl/\/\hss}}
  \nc{\bea}{\begin{eqnarray}}
  \nc{\eea}{\end{eqnarray}}
  \newcommand{\bq}{\begin{eqnarray*}}
  \newcommand{\eq}{\end{eqnarray*}}
  \nc{\be}{\bea}
  \nc{\ee}{\eea}
  \nc{\benu}{\begin{enumerate}}
  \nc{\eenu}{\end{enumerate}}

  \nc{\trac}[2]{{\textstyle\frac{#1}{#2}}}

  \nc{\ex}[1]{\mbox{e}^{\,\textstyle#1}}

  \nc{\mat}[4]{\left(\begin{array}{cc}#1&#2\\#3&#4\end{array}\right)}

  \nc{\som}[9]{\left(\begin{array}{ccc}#1&#2&#3\\#4&#5&#6\\#7&#8&#9%
  \end{array}\right)}

  \nc{\subs}[1]{{\vspace*{0.5cm}}%
  {\noindent\underline{#1}}{\addcontentsline{toc}{subsection}{#1}}%
  {\vspace*{0.3cm}}}

\begin{document}
\begin{titlepage}
\begin{flushright}
\end{flushright}
  \global\parskip=4pt

 \makeatletter

\vskip .5in
\vskip .5in
 {\bf
Nonlinear spinor field equations in gravitational theory: 
spherical symmetric soliton-like solutions }

 {\bf V. Adanhounme$^a$, A. Adomou$^a$,  F. P. Codo$^a$ and M. N. Hounkonnou$^{a}$\footnote{ Correspondence author:
norbert.hounkonnou@cipma.uac.bj (with copy to hounkonou@yahoo.fr)}}\\
 {\em $^a$University
of Abomey-Calavi}, \\{\em International Chair in Mathematical Physics and
Applications}, \\({\em ICMPA-UNESCO Chair}),  {\em 072 B.P. 50 Cotonou, Republic of
Benin}

\vspace{1.0cm}

\today
\begin{center}
\begin{abstract}
 This paper deals with an extension of a previous work [{{\it Gravitation \& Cosmology}},
  Vol. { 4}, 1998, pp 107--113] to
exact spherical symmetric solutions to the spinor field equations
with nonlinear terms which are arbitrary functions of
  $S=\overline{\psi }\psi $, taking into account their own gravitational
   field. Equations with power and polynomial nonlinearities are studied
    in detail. It is shown that the initial set of the Einstein and spinor
     field equations with  a power nonlinearity has regular solutions
     with  spinor field localized energy and charge densities. The
     total energy and charge are finite. Besides,
     exact solutions, including soliton - like
solutions, to the spinor field equations are also obtained in flat
space-time.
\end{abstract}
\end{center}

\textbf{Key-words}: Lagrangian, static spherical symmetric metric,
field equations, Einstein equations, Dirac equation, energy-momentum
tensor,  charge density, current vector, soliton-like solution.


 \end{titlepage}
\setcounter{footnote}{0}
 \makeatother

 \section{ Introduction }
The unification of quantum mechanics and general relativity into a
theory of quantum gravity remains a hard (as yet) unsolved problem
and  physical phenomena requiring both general relativity and
quantum theory for their description cannot be possibly completely
 understood. Such a challenge stimulates intense research activities  in
 various field-theoretical models with full non-perturbative account
 of gravity. Among all these activities, the investigations of
 solitons in these theories, with a special emphasis on flat space
 theories, attracted  a particular importance due to
 their properties. Indeed, the soliton sector in the flat space
 gauge theories is quite well understood, the most notable example
 being the t'Hooft-Polyakov magnetic monopole. For a review on some
 recent progress in the investigation of solitons and black holes in
 non-Abelian gauge theories coupled to gravity, see \cite{gal} and
 references therein. However, as is well known, the marriage of
 gravity and relativity leads to a curved space-time whose geometry
 is dynamical and is governed by the energy-matter distribution
 within it, a framework within which the gravitational interaction
 is the physical manifestation of any curvature in space and in
 space-time. The most fascinating offsprings of this union are
 undoubtedly, on the one hand, the cosmological theory of the
 history of our universe from its birth to its ultimate demise if
 ever, and on the other hand, the prediction for regions of
 space-time to be so much curled up by their energy-matter content
 that even light can no longer escape from such black holes.

 On the other hand, the marriage of relativity and quantum theory
 leads naturally to the quantum field theory description of the
 elementary particles and their interactions, at the most intimate
 presently accessible scales of space and energy, a fact made
 manifest by the value of the product $\hbar c\simeq 197$
 Mev.fm. In fact, one offspring of this second union is the
 unification of matter and radiation, namely of particles with their
 corpuscular propagating properties and fields with their wavelike
 propagating properties. Particles, characterized through their
 energy, momentum and spin values in correspondence with the
 Poincar\'e symmetries of Minkowski space-time in the absence of
 gravity, are nothing but the relativistic energy-momentum quanta of
 a field, thereby implying a tremendous economy in the description
 of the physical universe, accounting for instance at once in terms
 of a single field filling all of space-time for the
 indistinguishability of identical particles and their statistics.
 Furthermore, quantum relativistic interactions are then understood
 simply as couplings between the various quantum fields locally in
 space-time, which translate in terms of particles as diverse
 exchanges of the associated quanta. Such a picture lends itself
 most ideally to a perturbative understanding of the fundamental
 interactions, which has proved to be so powerful beginning with
 quantum electrodynamics, up to the modern Standard Model of the
 strong and electroweak interactions. For more explanation on these
 profound concepts, quantum theory and relativity, which have
 culminated into relativistic space-time geometry and quantum gauge
 theory as the principles for gravity and the three other known
 fundamental interactions, see notes \cite{govaerts} on {\it The quantum
 geometer's universe: particles, interactions and topology}
 delivered in 2001 by Govaerts at the Second International Workshop on
 Contemporary Problems in Mathematical Physics.

All these activities, diverse and complementary, made in this field
\cite{gal}-\cite{bif}, are also mainly motivated by the wide roles
of Einstein and Dirac equations in modern physics, for example, for
investigating the spin particle and for the necessity of analysis of
synchrotronic radiation \cite{bi8} . To this purpose, many systems
have been subjects of considerable interest and studies. The
pioneering investigation could be the work by Drill and Wheeler in
1957 \cite{1}, who considered the Dirac equation in a central
gravitational field associated with a diagonal metric. Using a
normal diagonal tetrad, these authors constructed the generalized
angular momentum operator separating the variables in the Dirac
equation. Later, in a remarkable paper, appeared in 1987 \cite{sa},
entitled "{\it Criteria of separability of variables in the Dirac
equation in gravitational fields}",  Shishkin and Andrushkevich
provided the necessary and sufficient conditions, based on rigorous
theorems, for separability of the variables for a diagonal tetrad
gauge, and deduced the operators that determine the dependence of
the wave function on the separated variables. In the same year,
Barut and Duru \cite{bd} gave exact solutions of the Dirac equation
in spatially flat Robertson-Walker space-times for models of
expanding universes and discussed the current decomposition.
Henceforth the investigations go into diverse directions,
considering various classes of models including different metrics,
the general class of which is investigated by Hounkonnou and Mendy
in 1999 \cite{hounkonnou}.
 Thus, for example, the usual
Friedman-Lema\^itre-Roberston-Walker homogeneous and isotropic
metric of standard cosmology belongs to this general class of metrics
(whether in Cartesian or spherical coordinates), which also includes
general classes of Kantowski-Sachs metrics for anisotropic
cosmologies as well as some examples of metrics used in models for
stellar gravitational collapse \cite{bif}. It may be worth pointing
out that {\it a priori}, this class of metrics solves Einstein's
equations for specific distributions of energy-momentum of matter in
space-time, in the presence of which the study of the quantized
Dirac field may be of interest. Such an avenue could be pursued. For
details, see \cite{hounkonnou} and references therein.

Moreover, it is also worthy of attention a previous study, which
will be referred to {\it Part I} of the present work, where Adomou
and Shikin \cite{6} have obtained exact plane-symmetric solutions to
the spinor field equations with nonlinear terms which are arbitrary
functions of $S=\overline{\psi }\psi $, taking into account their
own gravitational field. They  have studied in detail equations with
power and polynomial nonlinearities. They have shown that the
initial set of the Einstein and spinor field equations with a
power-law nonlinearity has regular solutions with a localized energy
density of the spinor field only in
 the case of zero mass parameter in the spinor field, with a
 negative energy for the soliton-like configuration. They have also proved
 that the spinor field equation with a polynomial nonlinearity has a
 regular solution with positive energy. Their study
  has  come out onto the non existence of  soliton-like solutions  in the flat space-time.

The present work, considered  as {\it Part II} of all these
investigated initiated in \cite{6}, aims at extending the results to
exact spherical symmetric solutions. Here also  equations with power
and polynomial nonlinearities are thoroughly scrutinized.

The paper is organized as follows. Section 2 addresses the model
with fundamental equations. We consider a self-consistent system to
obtain spherical-symmetric solutions, taking into account the own
 gravitational field of particles.  Section 3 deals with main
 results and their discussion; the solutions of the
  Einstein and nonlinear spinor field equations are derived. Besides,
the regularity
  properties of the obtained solutions as well as the
 asymptotic behavior of the energy and charge densities are
 studied. Concluding remarks are outlined in section 4.

 \section{Model and fundamental equations}
We consider the Lagrangian of the self-consistent
 system of spinor and gravitational fields in the form \cite{6}:

\begin{equation}
\label{eq1}
L=\frac{R}{2\kappa } +L_{sp}
\end{equation}

\begin{equation}
L_{sp} =\frac{i}{2}
 \left(\overline{\psi }\gamma ^{\mu }
 \nabla _{\mu } \psi -\nabla _{\mu }
  \overline{\psi }\gamma ^{\mu }
  \psi \right)-m\overline{\psi }\psi +L_{N}
 \end{equation}
where $R$ is the scalar curvature; $\kappa $ is Einstein's
gravitational constant and
 $L_{N} =F(S)$ is an arbitrary function depending
 on $S=\overline{\psi }\psi $.

Instead of the static plane-symmetric metric chosen in \cite{6}, in
the present analysis we opt for the static spherical symmetric
metric in
 the form:
 \begin{equation}
 \label{eq2}
 ds^{2} =e^{2\gamma } dt^{2} -e^{2\alpha }
 d\xi ^{2} -e^{2\beta } \left(d\theta ^{2} +\sin ^{2}
 \theta d\varphi ^{2} \right),
 \end{equation}
 $\alpha ,\, \beta ,\, \gamma $ being some functions
 depending only on $\xi =\frac{1}{r}$, where $r$ stands for the  radial
 component of the spherical symmetric metric, and satisfying the coordinate condition
 \begin{equation}
\label{eq3} \alpha =2\beta +\gamma.
\end{equation}
From the Lagrangian (\ref{eq1}), through the variational principle
and usual algebraic manipulations, one can readily deduce the
Einstein equations for the metric (\ref{eq2}) under the condition
(\ref{eq3}), the spinor field
 equations for the functions $\psi$, $\overline{\psi }$,
 and the components of the metric spinor field energy-momentum tensor,
  respectively,  in the form \cite{1}:
 \begin{eqnarray}
G^0_0&=&e^{-2\alpha } \left(2\beta "-\beta ^{'2}
-2\beta'\gamma'\right)-e^{-2\beta } = -\kappa T_{o}^{o}\label{g00} \\
G^1_1&=&e^{-2\alpha } \left(\beta ^{'2}
  +2\beta ^{'} \gamma ^{'} \right)-e^{-2\beta }
  =-\kappa T_{1}^{1} \label{g11} \\
  G^2_2&=&e^{-2\alpha }
   \left(\beta "+\gamma "-\beta ^{'2} -2\beta '
      \gamma '\right)=-\kappa T_{2}^{2}\label{g22}\\
      G^3_3&=&G^2_2\label{g33}\\
      T_{2}^{2} &=&T_{3}^{3}
 \end{eqnarray}
\begin{eqnarray}
i\gamma ^{\mu } \nabla _{\mu }
\psi -m\psi +L_{N}^{'} \psi &=&0 \label{eq25}\\
i\nabla _{\mu } \overline{\psi }\gamma ^{\mu }
     +m\overline{\psi }-L_{N}^{'} \overline{\psi }&=&0
\end{eqnarray}
\begin{equation}
T_{\mu \nu } =\frac{i}{4} \left(\overline{\psi }
\gamma _{\mu } \nabla _{\nu } \psi +\overline{\psi }
\gamma _{\nu } \nabla _{\mu } \psi -\nabla _{\mu }
 \overline{\psi }\gamma _{\nu } \psi -\nabla _{\nu }
  \overline{\psi }\gamma _{\mu } \psi \right)-g_{\mu \nu }
  L_{sp}
\end{equation}
where $\nabla _{\mu }$ is the covariant spinor
 derivative \cite{1}:
 $\nabla _{\mu } \psi =\frac{\partial \psi }{\partial \xi ^{\mu } }
 -\Gamma _{\mu } \psi $;
 $\Gamma _{\mu } \left(\xi \right)$ are the spinor affine connection matrices.
To define the matrices $\gamma ^{\mu } \left(\xi \right)$, let us
use the equalities
\begin{equation}
 \label{eq8}
 g_{\mu \nu } \left(\xi \right)=
e_{\mu }^{\left(a\right)} \left(\xi \right)e_{\nu }^{\left(b\right)}
\left(\xi \right)\eta _{ab} \, \, ;\, \, \gamma _{\mu } \left(\xi \right)
=e_{\mu }^{\left(a\right)} \left(\xi \right)\overline{\gamma _{a}}
\end{equation}
where $\eta _{ab} =diag\left(1,-1,-1,-1\right)$; $\overline{\gamma
_{a} }$  are the Dirac's matrices in flat space-time;
 $e_{\mu }^{\left(a\right)}
\left(\xi \right)$ are tetradic 4-vectors. Then we get:
\begin{equation}
\label{eq9}
 \gamma ^{o} =e^{-\gamma } \overline{\gamma }^{o} \, \, \, ;\, \, \,
  \gamma ^{1} =e^{-\alpha } \overline{\gamma }^{1}
  \, \, \, ;\, \, \, \, \gamma ^{2} =e^{-\beta }
   \overline{\gamma }^{2} \, \, \, ;\, \, \, \,
   \gamma ^{3} =\frac{e^{-\beta } \overline{\gamma }^{3} }{\sin \theta
   }.
     \end{equation}
 The matrices
 $\Gamma _{\mu } \left(\xi \right)$
 are then determined as follows:
\begin{eqnarray}
\Gamma _{\mu }& =&\frac{1}{4}
g_{\rho \delta } \left(\partial _{\mu }
e_{\sigma }^{\left(b\right)} .e_{\left(b\right)}^{\rho }
-\Gamma _{\mu \sigma }^{\rho } \right)\gamma ^{\delta } \gamma ^{\sigma }\label{eqn15};\\
\Gamma _{0} &=& -\frac{1}{2} \overline{\gamma }^{0} \overline{\gamma }^{1}
 e^{-2\beta } \gamma ^{'} \, \, \, ;\, \, \, \,
 \Gamma _{1} =0\, \, \, ;\, \, \, \,
 \Gamma _{2} =\frac{1}{2} \overline{\gamma }^{2} \overline{\gamma }^{1}
  e^{-\gamma -\beta } \beta ^{'} \label{eqn16}\\
  \Gamma _{3} &=& \frac{1}{2} \left(\overline{\gamma }^{3}
\overline{\gamma }^{1} e^{-\beta -\gamma } \beta ^{'} \sin \theta
+\overline{\gamma }^{3} \overline{\gamma }^{2} \cos \theta\right)\label{eqn17}.
\end{eqnarray}
The matrices $\overline{\gamma }^{a} $ are chosen as in [3]. Using
the spinor field equations, we can rewrite $L_{sp}$ in the form
\begin{equation}
\label{eq11}
L_{sp} =-\frac{1}{2} \left(\overline{\psi }
\frac{\partial L_{N} }{\partial \overline{\psi }} +
\frac{\partial L_{N} }{\partial \psi } \psi \right)+
L_{N} =-SL_{N}^{'} +L_{N},
\end{equation}
with the spinor $$
\psi =\left(\begin{array}{l} {V_{1} }
 \\ {V_{2} } \\ {V_{3} } \\ {V_{4} } \end{array}\right).
 $$
 Taking into account (\ref{eq11}),
 let us write explicitly the nonzero components of
  the tensor $T_{\nu }^{\mu } $:

\begin{equation}
\label{eq12}
T_{0}^{0} =T_{2}^{2} =T_{3}^{3}
=-L_{sp} =SL_{N}^{'} -L_{N}
  \end{equation}
 setting the condition $\overline{V }_1V_4 + \overline{V }_2V_3= \overline{V }_3V_2 +\overline{V }_4V_1 $,
  \begin{equation}
\label{eq13}
T_{1}^{1} =\frac{i}{2} \left(\overline{\psi }\gamma ^{1}
\nabla _{1} \psi -\nabla _{1} \overline{\psi }\gamma ^{1}
 \psi \right)+SL_{N}^{'} -L_{N}
  \end{equation}
  
  Using the obtained expressions for
 $\Gamma _{\mu } \left(\xi \right)$ in
 (\ref{eqn15}) - (\ref{eqn17}),
 we can expand  (\ref{eq25}) as
  \begin{equation}
 \label{eq14}
  ie^{-\alpha } \overline{\gamma }^{1}
  \left[\partial _{\xi } +\frac{1}{2}
  \alpha ^{'} \right]\psi +\frac{i}{2} e^{-\beta }
   \overline{\gamma }^{2} \psi\cot\theta -m\psi
   +L_{N}^{'} \psi =0
  \end{equation}
    yielding the following set of equations:
\begin{eqnarray}
V_{4}^{'} +\frac{1}{2} \alpha ^{'} V_{4} -\frac{i}{2}
e^{\alpha -\beta } V_{4} \cot\theta -ie^{\alpha }
\left(L_{N}^{'} -m\right)V_{1} &=&0\label{eq15a}
\\
V_{3}^{'} +\frac{1}{2} \alpha ^{'} V_{3}
 +\frac{i}{2} e^{\alpha -\beta } V_{3} \cot\theta
  -ie^{\alpha } \left(L_{N}^{'} -m\right)V_{2} &=&0 \label{eq15b}\\
-V_{2}^{'} -\frac{1}{2} \alpha ^{'} V_{2} +\frac{i}{2}
e^{\alpha -\beta } V_{2} \cot\theta -ie^{\alpha }
\left(L_{N}^{'} -m\right)V_{3} &=&0\label{eq15c}\\
-V_{1}^{'} -\frac{1}{2} \alpha ^{'} V_{1} -\frac{i}{2}
 e^{\alpha -\beta } V_{1} \cot\theta -ie^{\alpha }
 \left(L_{N}^{'} -m\right)V_{4} &=&0. \label{eq15d}
 \end{eqnarray}

 \section{ Results and discussion}

From the set of equations (\ref{eq15a})-(\ref{eq15d}), we
 infer that  the invariant function
$$S=\overline{\psi }\psi =V_{1}^{*} V_{1} +V_{2}^{*}
V_{2} - V_{3}^{*} V_{3} -V_{4}^{*} V_{4} $$ satisfies a first order
differential equation:
\begin{equation}
\label{eq31} \frac{dS}{d\xi } +\alpha ^{'}S
 =0
\end{equation}\label{invariant}
giving the evident solution
 \begin{equation}S=Ce^{-\alpha(\xi)},\end{equation}
 $C$ being a  constant.
Combining the spinor field equation  (\ref{eq14}) with its
  conjugate expression results   the following expression
  for (\ref{eq13}):
\begin{equation}
\label{eq33} T_{1}^{1} =mS-L_{N}.
  \end{equation}
  The difference $\left(\begin{array}{l} {0} \\ {0}
\end{array}\right)-\left(\begin{array}{l} {2} \\ {2}
 \end{array}\right)$ of the Einstein equations with
 (\ref{eq12}) leads to
\begin{equation}
\label{eq34}
\beta ^{"} -\gamma ^{"}
=e^{2\beta +2\gamma }
\end{equation}
which can be transformed into a Liouville equation (see [7], page
30) to produce the solutions:
\begin{eqnarray}
 \beta \left(\xi \right)&=&\frac{A}{4}
 \left(1+\frac{2}{G} \right)\ln \frac{A}{GT^{2}
 \left(h, \xi +\xi _{1} \right)}= \left(1+\frac{2}{G} \right)  \gamma \left(\xi \right)\label{eq35}\\
 \gamma \left(\xi \right)&=&
 \frac{A}{4} \ln \frac{A}{GT^{2}
 \left(h\, ,\, \xi +\xi _{1} \right)}
  \label{eq36}
 \end{eqnarray}
 where the quantity $A$ is expressed in terms of the  Newton's gravitational constant $G$ as:
  $$A=\frac{G}{G+1}.$$
 \begin{equation}
 T\left(h\, ,\, \xi +\xi _{1} \right)=
\left\{\begin{array}{l} {\frac{1}{h}
\sinh \left[h\left(\xi +\xi _{1} \right)\right], h>0}
 \\
  {\xi +\xi _{1}, h=0} \\
  {\frac{1}{h} \sin \left[h\left(\xi +\xi _{1}
  \right)\right], h<0}, \end{array}\right.
  \end{equation}
  $h$ being an integration constant and $\xi _{1}$
another non zero integration constant. Taking into account
(\ref{eq35}) and (\ref{eq36}), we get from (\ref{eq3})
 the following relations:
\begin{equation}
\label{eq37}
\alpha \left(\xi \right)=\frac{A}{2}
 \left(\frac{3}{2} +\frac{2}{G} \right)
 \ln \frac{A}{GT^{2} \left(h,\, \, \xi +\xi _{1} \right)}
\end{equation}
and
\begin{equation}
\beta \left(\xi \right)=\frac{2+G}{4+3G}
 \alpha \left(\xi \right)\, \, ;\, \,
  \gamma \left(\xi \right)=\frac{G}{4+3G}
  \, \alpha \left(\xi \right).
  \end{equation}
  Substituting (\ref{eq37}) into (\ref{g11}),
 we obtain the Einstein equation
  $\left(\begin{array}{l}
  {1} \\ {1} \end{array}\right)$
  in the form
\begin{equation}
\label{eq38}
 \alpha ^{'2} =\frac{\left(4+3G\right)^{2} }
 {3G^{2} +8G+4} e^{2\alpha } \left[e^{\frac{-2G-4}{4+3G}
 \, \alpha } -\kappa \left(mS-L_{N} \right)\right].
 \end{equation}
Since $\alpha ^{'} =-\frac{1}{S} \frac{dS}{d\xi } \, $with the
invariant $S=Ce^{-\alpha }$, from (\ref{eq38}), we get:
\begin{equation}
\label{eq39}
\frac{dS}{d\xi } =\pm \frac{\left(4+3G\right)S}
{\sqrt{3G^{2} +8G+4} } \left(\frac{C}{S} \right)
\sqrt{\left(\frac{S}{C} \right)^{\frac{4+2G}{4+3G} }
 -\kappa \left(mS-L_{N} \right)}.
  \end{equation}
 With the knowledge of  $\beta \left(\xi \right),\, \,
   \gamma \left(\xi \right)$ and $\alpha
  \left(\xi \right)$ from the relations (\ref{eq35}), (\ref{eq36}) and
 (\ref{eq37}), respectively,  the invariant $S\left(\xi \right)$ as well as
 the solutions of the Einstein equations
  can be  completely determined. Furthermore,
considering the concrete expression of the invariant $S\left(\xi
\right)$, namely $S\left(\xi \right)= Ce^{-\alpha \left(\xi
\right)}$,  we can establish
 the regularity properties of the obtained solutions.
 Studying the distribution of the energy per unit invariant
  volume $T_{0}^{0} \sqrt{^{-3} g} $, we can also deduce
   their localization properties.

We can get a concrete form of the functions $V_{\rho }
 \left(\xi \right)$ by solving equations (\ref{eq15a})-(\ref{eq15d})
  in a more compact form if we pass to the functions
  $W_{\rho } \left(\xi \right)=e^{\frac{1}{2}
  \alpha \left(\xi \right)} V_{\rho } \left(\xi \right),\,
  \rho =1,2,3,4:$
  \begin{eqnarray}
W_{4}^{'} -\frac{i}{2} e^{\alpha -\beta } W_{4} \cot \theta
-ie^{\alpha } \left(-m+L_{N}^{'}
\right)W_{1} &=&0 \label{eq310a}\\
W_{3}^{'} +\frac{i}{2} e^{\alpha -\beta } W_{3}
 \cot \theta -ie^{\alpha } \left(-m+L_{N}^{'}
 \right)W_{2} &=&0\\
   W_{2}^{'} -\frac{i}{2} e^{\alpha -\beta } W_{2}
 \cot \theta +ie^{\alpha } \left(-m+L_{N}^{'} \right)W_{3} &=&0 \\
W_{1}^{'} +\frac{i}{2} e^{\alpha -\beta }
W_{1} \cot \theta +ie^{\alpha } \left(-m + L_{N}^{'}
 \right)W_{4} &=&0   \label{eq310b}
 \end{eqnarray}
 where
 \begin{eqnarray}
W_{\rho }^{'} =\left(V_{\rho }^{'} +\frac{1}{2} \alpha ^{'} V_{\rho
} \right)e^{\frac{1}{2} \alpha }.
\end{eqnarray}
Re-express eqs. (\ref{eq310a})-(\ref{eq310b})
under forms depending on  functions of the argument
$S\left(\xi \right)$, i.e. 
$ U_{\rho } \left(S\right)=W_{\rho }
\left(\xi \right)$, $
 S\left(\xi \right)=
 Ce^{-\alpha \left(\xi \right)}.$
 Then we get for the functions $U_{\rho } \left(S\right)$
the following set of equations:
\begin{eqnarray}
\frac{dU_{4} }{dS} -iB\left(S\right)U_{4} -iQ\left(S\right)U_{1} &=&0
\label{eq311a}\\
\frac{dU_{3} }{dS} +iB\left(S\right)U_{3} -iQ\left(S\right)U_{2} &=&0
\label{eq311b}\\
\frac{dU_{2} }{dS} -iB\left(S\right)U_{2} + iQ\left(S\right)U_{3} &=&0
\label{eq311c}\\
\frac{dU_{1} }{dS} +iB\left(S\right)U_{1} + iQ\left(S\right)U_{4} &=&0
\label{eq311d}
\end{eqnarray}
where
\begin{equation}
 \label{eq312}
 B\left(S\right)=\frac{1}{2}
  \frac{\left(\frac{C}{S} \right)^{\frac{2+2G}{4+3G} }
  \cot \theta }{\frac{dS}{d\xi }};
\quad
    Q\left(S\right)=\frac{\left(\frac{C}{S} \right)
    \left(-m+L_{N}^{'} \right)}{\frac{dS}{d\xi } }
 \end{equation}
with $\frac{dS}{d\xi } $ determined by (\ref{eq39}).

 Differentiating now
 eqs. (\ref{eq311a})-(\ref{eq311d}) and substituting
 eqs. (\ref{eq311d}) and (\ref{eq311a}) into the result, we obtain
   second-order differential equations obeyed by the functions $U_{4}
   \left(S\right)$ and $U_{1} \left(S\right)$:

\begin{equation}
\label{eq313}
 U_{4}^{\prime\prime}
-\frac{Q^{'} \left(S\right)}{Q\left(S\right)}
U_{4}^{'} +\left[B^{2} \left(S\right)-Q^{2}
\left(S\right)+i\frac{B\left(S\right)Q^{'} \left(S\right)
-Q\left(S\right)B^{'} \left(S\right)}{Q\left(S\right)} \right]U_{4} =0
\end{equation}

\begin{equation}
\label{eq314} U_{1}^{\prime\prime}
 -\frac{Q^{'} \left(S\right)}{Q\left(S\right)}
 U_{1}^{'} +\left[B^{2} \left(S\right)-Q^{2}
 \left(S\right)+i\frac{Q\left(S\right)B^{'}
 \left(S\right)-B\left(S\right)Q^{'}
 \left(S\right)}{Q\left(S\right)} \right]U_{1} =0\,
  \end{equation}

Summing (\ref{eq313}) and (\ref{eq314}) and setting $U=U_{1} +U_{4}$
afford the differential equation:
\begin{equation}
\label{eq315}
 U^{\prime\prime} -\frac{Q^{'}
 \left(S\right)}{Q\left(S\right)} U^{'}
 +\left[B^{2} \left(S\right)-Q^{2}
 \left(S\right)\right]U=0,
 \end{equation}
which, under the condition $B^{2} \left(S\right)=\left(1-\varepsilon
 \right)Q^{2} \left(S\right),$ with
 $0<\varepsilon \le 1,\, \, $
  yields
  the solution
  \begin{equation}
  \label{eq316}
   U_{1} +U_{4}
   =\alpha _{0} \, \cosh N_{1} \left(S\right),\,  \alpha _{0} =const,
   \end{equation}
where  $N_{1} \left(S\right)=\sqrt{\varepsilon } \int
Q\left(S\right) dS+R_{1}
 \, \, ;\, \, \, R_{1} =const$.
Substracting eqs. (\ref{eq311a}) and (\ref{eq311d}) and taking into
account  (\ref{eq316}), we obtain
  \begin{equation}
 \label{eq318}
 U_{1} -U_{4}
  =-i\alpha _{0} \frac{\sqrt{1-\varepsilon } +1}{\sqrt{\varepsilon } }
  \sinh N_{1} \left(S\right).
  \end{equation}

It then follows, from the  equations (\ref{eq316}) and
(\ref{eq318}),
 that
\begin{equation}
\label{eq319}
 U_{1}
\left(S\right)=\alpha _{1} \left[\cosh N_{1}
\left(S\right)-i\frac{1+\sqrt{1-\varepsilon } } {\sqrt{\varepsilon }
} \sinh N_{1} \left(S\right)\right]
\end{equation}
and
\begin{equation}
 \label{eq320}
  U_{4} \left(S\right)=
\alpha _{1} \left[\cosh N_{1} \left(S\right)+
i\frac{1+\sqrt{1-\varepsilon } }{\sqrt{\varepsilon } }
 \sinh N_{1} \left(S\right)\right]
 \end{equation}
with $\alpha _{1}=\frac{\alpha _{0} }{2}$.

Analogously operating on eqs. (\ref{eq311b}) and (\ref{eq311c}),
 we arrive at
\begin{equation}
\label{eq321}
U_{2} \left(S\right)=\alpha _{2}
 \left[\sinh N_{2} \left(S\right)-
 i\frac{-1+\sqrt{1-\varepsilon } }{\sqrt{\varepsilon } }
  \cosh N_{2} \left(S\right)\right]
   \end{equation}
and
\begin{equation}
\label{eq322}
 U_{3} \left(S\right)=\alpha _{2}
 \left[\sinh N_{2} \left(S\right)+i\frac{-1+\sqrt{1-\varepsilon } }
 {\sqrt{\varepsilon } } \cosh N_{2} \left(S\right)\right],
   \end{equation}
with $\alpha _{2}= const$,
\begin{equation}
\label{eq323}
 N_{2} \left(S\right)=
 -\sqrt{\varepsilon } \int Q\left(S\right)
 dS+R_{2} \, \, ,\, \, R_{2}= const.
\end{equation}

 As mentioned in \cite{6}, it is worth considering a self-consistent
 solution to the linear spinor field equation
 (Dirac's equation), in view of its comparison with solutions to nonlinear
  spinor equations
 and of a better insight of the role of nonlinear terms
 in the nonlinear field equations in the formation of regular
 localized soliton-like solutions.  For this purpose, $L_{N} =0$ and we
  have from (\ref{eq14}):
\begin{equation}
 \label{invariant}
  ie^{-\alpha } \overline{\gamma }^{1}
  \left[\partial _{\xi } +\frac{1}{2}
  \alpha ^{'} \right]\psi +\frac{i}{2}
  e^{-\beta } \overline{\gamma }^{2}
  \psi \cot \theta -m\psi =0.
  \end{equation}
In this case, the relation ( \ref{invariant}) giving
 $S\left(\xi \right)$ becomes:
\begin{equation}
\label{GrindEQ__3_25_}
S\left(\xi \right)=C\exp
\left\{-\frac{A}{2}
 \left(\frac{3}{2} +\frac{2}{G}
 \right)\ln \frac{A}{GT^{2}
 \left(h,\xi +\xi _{1} \right)} \right\}.
 \end{equation}
From \eqref{eq35}, \eqref{eq36} and \eqref{eq37},
 we get:
\begin{equation}
 \label{GrindEQ__3_26_}
 e^{2\gamma \left(\xi \right)}
 =\exp \left\{\frac{A}{2} \ln
 \frac{A}{GT^{2} \left(h,\xi +\xi _{1}
 \right)}
  \right\}
  \end{equation}
\begin{equation}
 \label{GrindEQ__3_27_}
 e^{2\beta \left(\xi \right)}
 =\exp \left\{\frac{A}{2}
 \left(1+\frac{2}{G} \right)
 \ln \frac{A}{GT^{2}
 \left(h,\xi +\xi _{1} \right)} \right\}
  \end{equation}
\begin{equation}
 \label{GrindEQ__3_28_}
 e^{2\alpha \left(\xi \right)}
 =\exp \left\{A\left(\frac{3}{2} +\frac{2}{G}
 \right)\ln \frac{A}{GT^{2} \left(h,\xi +\xi _{1} \right)}
  \right\}
  \end{equation}
showing that
 the invariant $S$ and the functions $g_{00} =e^{2\gamma }$,
 $ g_{11} =-e^{2\alpha }$
 $ g_{22} =-e^{2\beta }$
$g_{33} =-e^{2\beta } \sin ^{2} \theta \, $are regular. In the case
under consideration we have $T_{0}^{0} \left(\xi \right)=0$,  i.e.
the energy density is localized.

Using \eqref{eq39}, \eqref{eq312} and \eqref{eq323}, we get:
\begin{equation}
\label{GrindEQ__3_29_}
 N_{1,2}
\left(S\right)=\frac{\sqrt{C^{a} \varepsilon \left(3G^{2}
+8G+4\right)} }{4+3G} \int \frac{-m+L_{N}^{'} }{S\sqrt{S^{a} -C^{a}
\kappa \left(mS-L_{N} \right)} } dS,
\end{equation}
with $a=\frac{4+2G}{4+3G} \cong 1$.

Let us find the explicit form of $V_{\rho } \left(\xi \right),\, \,
\rho =1,2,3,4$. To this end, we retrieve the expressions of $N_{1}
 \left(S\right)\, \, \, {\rm and}\, \, \, \, N_{2}
  \left(S\right)$ from \eqref{GrindEQ__3_29_}, knowing that
  $L_{N} =0$. Without loss of generality, let us set $a=1$. Then,
\begin{equation}
 \label{GrindEQ__3_30_} N_{1,2}
  \left(S\right)=\frac{2m\sqrt{\varepsilon C\left(3G^{2}
  +8G+4\right)} }{\left(4+3G\right)\sqrt{\left(1-C\kappa m\right)S} }
   +R_{1,2}.
   \end{equation}
Substituting $S\left(\xi \right)$ from
 \eqref{GrindEQ__3_25_} into \eqref{GrindEQ__3_30_},
  we get
\begin{eqnarray}
\label{GrindEQ__3_31_} N_{1,2} \left(\xi
\right)=&&\frac{2m\sqrt{\varepsilon \left(3G^{2} +8G+4\right)} }
{\left(4+3G\right)\sqrt{1-C\kappa m} }\nonumber\\
&&\times \exp \left\{\frac{A}{4} \left(\frac{3}{2} +\frac{2}{G}
\right)\ln \frac{A}{GT^{2} \left(h,\xi +\xi _{1} \right)}
\right\}+R_{1,2},
 \end{eqnarray}
with $R_{1,2} =const$.

We then replace the expressions of $N_{1} \left(\xi \right)$
 and $N_{2} \left(\xi \right)$ from
 \eqref{GrindEQ__3_31_} into \eqref{eq319} -
 \eqref{eq322} and get  an explicit form of
 $U_{\rho } \left(\xi \right),\,  \rho =1,2,3,4$,
 and subsequently the expressions  of $V_{\rho }
 \left(\xi \right)=U_{\rho } \left(\xi \right)
 e^{-\frac{1}{2} \alpha \left(\xi \right)}$:
\begin{eqnarray}
\label{GrindEQ__3_32_}
V_{1} \left(\xi \right)=\alpha _{1}
\exp \left\{-\frac{A}{4}
\left(\frac{3}{2} +\frac{2}{G} \right)
\ln \frac{A}{GT^{2} \left(h,\xi +\xi _{1}
 \right)} \right\}\cr
 \qquad \times \left[\cosh N_{1}
 \left(\xi \right)-i\frac{1+\sqrt{1-\varepsilon } }{\sqrt{\varepsilon } }
 \sinh N_{1} \left(\xi \right)\right]
  \end{eqnarray}

\begin{eqnarray}
\label{GrindEQ__3_33_}
 V_{2} \left(\xi \right)
 =\alpha _{2} \exp \left\{-\frac{A}{4} \left(\frac{3}{2} +\frac{2}{G}
  \right)\ln \frac{A}{GT^{2} \left(h,\xi +\xi _{1} \right)}
   \right\}\cr
 \qquad\times \left[\sinh N_{2} \left(\xi \right)-i
   \frac{\sqrt{1-\varepsilon } -1}{\sqrt{\varepsilon } } \cosh N_{2}
    \left(\xi \right)\right]
     \end{eqnarray}

\begin{eqnarray}
\label{GrindEQ__3_34_}
V_{3} \left(\xi \right)=
\alpha _{2} \exp \left\{-\frac{A}{4}
 \left(\frac{3}{2} +\frac{2}{G} \right)
 \ln \frac{A}{GT^{2} \left(h,\xi +\xi _{1}
  \right)} \right\}\cr
 \qquad\times \left[\sinh N_{2}
  \left(\xi \right)+i\frac{\sqrt{1-\varepsilon }
   -1}{\sqrt{\varepsilon } } \cosh N_{2}
   \left(\xi \right)\right];
   \end{eqnarray}

\begin{eqnarray}
\label{GrindEQ__3_35_}
 V_{4} \left(\xi \right)
 =\alpha _{1} \exp \left
 \{-\frac{A}{4} \left(\frac{3}{2}
 +\frac{2}{G} \right)\ln \frac{A}{GT^{2}
  \left(h,\xi +\xi _{1} \right)} \right\}\cr
 \qquad\times
   \left[\cosh N_{1} \left(\xi \right)+i\frac{1+\sqrt{1-\varepsilon } }
   {\sqrt{\varepsilon } } \sinh N_{1} \left(\xi \right)\right].
     \end{eqnarray}
which represent nothing but  the regular localized soliton-like
solutions.

In the sequel, we deal with a concrete type of nonlinear spinor
field equations which have the virtue that $L_{N} =\lambda S^{n} $,
where $\lambda$ is a nonlinearity parameter, $n\ge 2$. It is
convenient to separately analyze the two cases  $n=2$ and $n>2$:
\begin{itemize}
\item $n=2$:
$L_{N} = \lambda S^{2} $ and we have the nonlinear spinor field
equation
\begin{equation}
\label{GrindEQ__3_36_} ie^{-\alpha } \overline{\gamma }^{1}
\left(\partial _{\xi } + \frac{1}{2} \alpha ^{'} \right)\psi
+\frac{i}{2} e^{-\beta } \overline{\gamma }^{2} \psi \cot \theta
-m\psi +2\lambda \left(\overline{\psi }\psi \right)\psi =0.
\end{equation}

The equalities \eqref{GrindEQ__3_25_}-\eqref{GrindEQ__3_28_}
remain valid. Let us find an explicit form of $V_{\rho }
 \left(\xi \right),\, \, \rho =1,2,3,4$. For that,
  we deduce from \eqref{GrindEQ__3_29_} the function $N_{1}
   \left(S\right)$ {\rm and}  $N_{2} \left(S\right)$:
\begin{eqnarray}
\label{GrindEQ__3_37_}
N_{1,2} \left(S\right)&=&\frac{\sqrt{C\varepsilon \left(3G^{2}
 +8G+4\right)} }{4+3G} \nonumber\\ &&\times \left\{\frac{2m}{\sqrt{C\kappa \lambda }
  +\sqrt{C\kappa \lambda S^{2} +\left(1-C\kappa m\right) S}} \right.
  \nonumber\\ &&\left. + 2\sqrt{\frac{\lambda }{C\kappa } } \ln
   \left[\frac{2C\kappa \lambda }{1-C\kappa m} S+1\right.\right.
    \nonumber\\ &&\left. \left.+
   \frac{2\sqrt{C\kappa \lambda } }{1-C\kappa m}
   \sqrt{C\kappa \lambda S^{2} +\left(1-C\kappa m\right)S} \right]
    \right\}+R_{1,2}\nonumber\\
    \end{eqnarray}
that we substitute into \eqref{eq319} - \eqref{eq322} to get  an
explicit expression of $U_{\rho } \left(\xi \right)$ and
subsequently   the initial functions $V_{\rho } \left(\xi \right)=\,
\, U_{\rho }
 \left(\xi \right)e^{-\frac{1}{2} \alpha \left(\xi \right)}$,
  $\rho =1,2,3,4$.

Let us compute the distribution of the spinor field energy density
per unit invariant volume$f\left(\xi \right)=T_{0}^{0} \left(\xi
\right)\sqrt{^{-3} g} $. From \eqref{eq12} and
\eqref{GrindEQ__3_25_} we have the following expression for
 $T_{0}^{0} \left(\xi \right)$:
\begin{equation}
\label{GrindEQ__3_38_}
T_{0}^{0} \left(\xi \right)=
\lambda S^{2} \left(\xi \right)=\lambda C^{2}
\exp \left\{-A\left(\frac{3}{2} +\frac{2}{G} \right)
\ln \frac{A}{GT^{2} \left(h,\xi +\xi _{1} \right)} \right\},
\end{equation}
permiting to write
\begin{eqnarray}f\left(\xi \right)&=&T_{0}^{0}
\left(\xi \right) e^{\alpha \left(\xi \right)+2\beta \left(\xi
\right)}
 \sin \theta \cr
&=&\lambda C^{2} \sin \theta \exp \left\{-\frac{A}{4}
 \ln \frac{A}{GT^{2} \left(h,\xi +\xi _{1}
  \right)} \right\}\label{GrindEQ__3_39_},
  \end{eqnarray}
inferring  that the quantities $g_{00}
 $, $g_{11} $, $g_{22} $ and $V_{\rho } $
 are regular and, from
 \eqref{GrindEQ__3_39_}, the total energy  $E=\int _{0}^{\xi _{c} }T_{0}^{0}
  \left(\xi \right)\sqrt{^{-3} g} d\xi $
  is  finite. Therefore,
  the equation \eqref{GrindEQ__3_36_} possesses a soliton-like solution.
\item $n>2$: $L_{N} =\lambda S^{n}$ and
 the energy density is
\begin{equation}
\label{GrindEQ__3_40_} T_{0}^{0} =\lambda \left(n-1\right)S^{n}.
 \end{equation}

From \eqref{GrindEQ__3_25_},
the distribution of the spinor field
 energy density per unit invariant volume takes the form
 $$
 {f\left(\xi \right)=T_{0}^{0} \left(\xi \right)\sqrt{^{-3} g} }$$
 i.e.
\begin{eqnarray}
\label{GrindEQ__3_41_} f\left(\xi \right)&=& \lambda
\left(n-1\right)C^{n}
 \sin \theta \exp \left\{\frac{A}{4G} \left[-n\left(4+3G\right)
 \right. \right. \cr && \left. \left.  +5G+8\right].\ln \frac{A}{GT^{2}
\left(h,\xi +\xi _{1} \right)}
 \right\}
 \end{eqnarray}
showing that
  the spinor field energy density per unit invariant
   volume $f$ is localized and the total energy
$E=\int _{0}^{\xi _{c} }T_{0}^{0}\left(\xi \right) \sqrt{^{-3} g} d\xi $  is
finite. To compute $V_{\rho } {\rm \; ,\; }\rho {\rm =1,2,3,4.}$, we
need  the functions $N_{1} \left(S\right)$ and
$N_{2}\left(S\right)$:
\begin{eqnarray}
 N_{1,2}
 \left(S\right)&=&
\frac{\sqrt{C\varepsilon (3G^{2} +8G+4)} }{4+3G}\cr
 &&\times \left\{\frac{2n}{C\kappa (n-2)} \sqrt{C\kappa
 \lambda S^{n-2} +(1-C\kappa m)\frac{1}{S} }\right.\cr
&&  +
 \frac{1}{\sqrt{C\kappa \lambda } } \left(\frac{C\kappa
  \lambda }{1-C\kappa m} \right)^{\frac{n}{2(n-1)} } }
{ \left[\frac{n}{n-2} \left(\frac{1}{C\kappa }
   -m\right)-m\right]\cr
  && \times  \left[B_{S} \left(\frac{n}{2(n-1)}
   ;1-\frac{1}{2(n-1)} \right)\right.\cr &&\left. -2\left(\frac{1-C\kappa m}
   {C\kappa \lambda S^{n-1} +1-C\kappa m} \right)^{\frac{n}{2(n-1)} }\right.\cr
   &&\left. \times \left(1+\frac{1-C\kappa m}{C\kappa \lambda S^{n-1} } \right)
    ^{\frac{1}{2(n-1)} } \right]\left. \right\}
    +R_{1,2},
\end{eqnarray}
    where
\begin{equation}
{B_{s}\left(\frac{n}{2(n-1)} ;1-\frac{1}{2(n-1)} \right)
=\int _{0}^{S}y^{\frac{n}{2(n-1)} -1}
 \left(1-y\right) ^{\frac{-1}{2(n-1)} }
 dy{\rm \; }}
 \end{equation}
  {\rm and\; }
  \begin{equation}
  y=
 \frac{1-C\kappa m}{C\kappa \lambda t^{n-1} +1-C\kappa m},
 \end{equation}
that we substitute into
 \eqref{eq319}-\eqref{eq322}
 to get  an explicit expression for  $U_{\rho }
 \left(\xi \right)$, and then we readily  compute
  the initial functions
 \begin{equation}
 \label{eq32}
 V_{\rho } \left(\xi \right)=U_{\rho }
  \left(\xi \right)e^{-\frac{1}{2} \alpha
  \left(\xi \right)}
  \end{equation} for
   $\rho =1,2,3,4$.
Using the solutions \eqref{eq319} - \eqref{eq322}, we deduce the
components of the spinor current vector $j^{\mu } =\overline{\psi
}\gamma ^{\mu } \psi$ as follows:
\begin{eqnarray}
\label{GrindEQ__3_42_}
 {j^{o}}
&=&2e^{-\gamma -\alpha } \left\{\alpha _{1}^{2} \left[\cosh^{2} N_{1}
\left(S\right)+\left(\frac{1+\sqrt{1-\varepsilon } }
{\sqrt{\varepsilon } } \right)^{2} \sinh^{2} N_{1}
\left(S\right)\right]\right.  \cr &&+ \left. \alpha _{2}^{2}
\left[\sinh^{2} N_{2} \left(S\right)+
\left(\frac{-1+\sqrt{1-\varepsilon } } {\sqrt{\varepsilon } }
\right)^{2} \cosh^{2} N_{2}
 \left(S\right)\right]\right\}\cr
  {j^{1}} &=&2e^{-2\alpha }
  \left\{\alpha _{1}^{2} \left[\cosh^{2} N_{1}
   \left(S\right)-\left(\frac{1+\sqrt{1-\varepsilon } }
   {\sqrt{\varepsilon } } \right)^{2} \sinh^{2} N_{1}
   \left(S\right)\right]\right.  \cr &&+\left. \alpha _{2}^{2}
     \left[\sinh^{2} N_{2} \left(S\right)-
     \left(\frac{-1+\sqrt{1-\varepsilon }
     }{\sqrt{\varepsilon } } \right)^{2} \cosh^{2} N_{2}
      \left(S\right)\right]\right\}
\cr j^{2} &=& 4e^{-\beta -\alpha } \left(\alpha _{1}^{2}
\frac{1+\sqrt{1-\varepsilon } } {\sqrt{\varepsilon } } \cosh N_{1}
\left(S\right).\sinh N_{1}
 \left(S\right)\right.\cr
 &&\left.-\alpha _{2}^{2} \frac{-1+\sqrt{1-\varepsilon } }
 {\sqrt{\varepsilon } } \sinh N_{2} \left(S\right)\cosh N_{2} \left(S\right)\right)
\cr
j^{3} &=&0.
 \end{eqnarray}

Since the configuration is static, only the component
 $j^{o}$ is nonzero. The constants in the solution of the spinor
  field equation are obtained from the equations $j^{1} =0$ and
 $ j^{2} =0$, thus giving
 $\alpha _{1} =\alpha _{2} ,
 \, \, \, N_{2} \left(S\right)=-N_{1}
 \left(S\right)$ and $\varepsilon =1$.
 The component $j^{o}$ defines the charge
 density of the spinor field whose  the chronometric
  invariant form  is characterized by:
\begin{equation}
 \label{GrindEQ__3_46_}
q=\left(j_{o} j^{o} \right)^{{\raise0.7ex\hbox{$ 1 $}\!\mathord{\left/
{\vphantom {1 2}} \right. \kern-\nulldelimiterspace}\!\lower0.7ex\hbox{$ 2 $}} }
 =4a^{2} e^{-\alpha } \cosh 2N\left(S\right)\,
 \end{equation}
where $a=\alpha _{1} =\alpha _{2} ,\, N\left(S\right)=N_{1}
\left(S\right)=-N_{2} \left(S\right),\,
 \varepsilon =1$. The total charge of the spinor field is:
 \begin{equation}
 \label{GrindEQ__3_47_}
 Q=\int _{o}^{\xi _{c} }q\sqrt{^{-3} g} d\xi,
 \end{equation}
$\xi _{c}$ being the center of the field configuration.

The relations \eqref{GrindEQ__3_25_}, \eqref{GrindEQ__3_31_},
 \eqref{GrindEQ__3_37_},
 \eqref{GrindEQ__3_46_} and \eqref{GrindEQ__3_47_} infer
 that the charge density of the spinor field is localized,
 and the total charge is a finite quantity, when
 $L_{N} =0$, or $ \lambda S^{2}$, or $\lambda S^{n},\, n>2$.
\end{itemize}
\section{ Concluding remarks}
In this paper, we  have obtained exact spherical symmetric solutions
to the spinor and gravitational field equations and  studied their
regularity properties as well as the localization properties of both
the energy and charge densities in different configurations, when
$L_{N} =0,\, \lambda S^{2}, \, {\rm and}\, \lambda S^{n} $.

In all these cases, the solutions are regular; the energy
 and  charge densities are localized. The total energy and
  charge of the spinor field are finite quantities.
The study of the set of all regular spherical solutions with a
possible criterion of their classification could deserve some
interest. Such investigation will be in the core of the forthcoming
paper.

\section*{Acknowlegments}
 This work is partially
supported by the Abdus Salam International Centre for Theoretical
Physics (ICTP, Trieste, Italy) through the
 OEA-ICMPA-\mbox{Prj-15}. The ICMPA is in partnership with
the Daniel Iagolnitzer Foundation (DIF), France.

  \end{document}